# Analysis of Non-Persistent CSMA Protocols with Exponential Backoff Scheduling

Pui King Wong, Dongjie Yin, and Tony T. Lee, *Fellow*, *IEEE*

*Abstract* — This paper studies the performance of Non-persistent CSMA/CA protocols with *K*-Exponential Backoff scheduling algorithms. A multi-queue single-server system is proposed to model multiple access networks. The input buffer of each access node is modeled as a Geo/G/1 queue, and the service time distribution of head-of-line packets is derived from the Markov chain of underlying scheduling algorithm. The main results include the complete analysis of the throughput and delay distribution, from which we obtained stable regions with respect to the throughput and bounded mean delay of the Geometric Retransmission and Exponential Backoff schemes. We show that the throughput stable region of Geometric Retransmission will vanish as the number of nodes $n \to \infty$; thus, it is inherently unstable for large *n*. In contrast to Geometric Retransmission, the throughput stable region of Exponential Backoff can be obtained for an infinite population. We found that the bounded mean delay region of Geometric Retransmission remains the same as its throughput stable region. Besides, the variance of service time of Exponential Backoff can be unbounded due to the capture effect; thus, its bounded delay region is only a sub-set of its throughput stable region. Analytical results presented in this paper are all verified by simulation.

## I. INTRODUCTION

Carrier Sense Multiple Access (CSMA) is one of the most widely deployed Media Access Control (MAC) protocols that allow multiple users to share a common transmission channel, such as 802.11 wireless LAN and Ethernet. Since Abramson's landmark paper on the ALOHA systems [1] embarked on the throughput analysis of multiple access channels, many researchers have followed his approach in studying other MAC protocols including CSMA. The analysis of CSMA can be traced back to Kleinrock and Tobagi's paper [2]. Under the assumption that there are infinite numbers of nodes and the aggregate traffic form a Poisson stream, they derived the throughput of various CSMA protocols by using an approach based on renewal theory. Although the characteristic equation of throughput can be easily obtained by using this elegant method, the ability to guarantee a stable throughput of the entire network and to bound the mean delay of packets are still open problems when analyzing the performance of CSMA protocols.

In the Non-Persistent Carrier-Sense Multiple Access (NP-CSMA) protocol, the node takes action after sensing the channel status. If the channel is in the idle state, the packet is sent immediately; otherwise, the packet is delayed for a random amount of time before sensing the channel again. Collisions may occur when two or more packets are simultaneously transmitted. Almost all follow-up research on the performance analysis of CSMA protocols [3]-[7] has been based on the network model originally proposed in [2]. For example, the characteristic equation for network throughput was re-derived in [3] by using a state-transition diagram of the transmission channel. The saturation throughput for NP-CSMA is deduced from the semi-Markovian property of the protocols in [4]. Some variants of slotted NP-CSMA with an adaptive array to improve throughput performance are proposed in [5].

Collisions are unavoidable in a shared environment, and a contention resolution algorithm is an essential mechanism in any MAC protocol to resolve contention among active nodes. However, previous analysis in [2]-[5] did not take the scheduling of collided packets into consideration.

Exponential Backoff algorithm is a widely adopted collision resolution scheme in most MAC protocols. It was shown in [6] that Exponential Backoff for MAC protocols was unstable for an infinite number of nodes. The authors of [7] claimed that if the arrival rate of the system is sufficiently small, the Exponential Backoff can still be stable. Despite these unfavorable observations, it is proven in [8] that a stable throughput can be achieved by the Exponential Backoff even if the number of nodes goes to infinity. Currently, a consensus on the stability issue of the Exponential Backoff scheme cannot be concluded from these controversial results.

A drastically different approach, based on a Markov chain model to analyze IEEE 802.11 with Exponential Backoff scheme, is proposed in [9], which provides extensive throughput analysis of the IEEE 802.11 with the Distributed Coordination Function (DCF) mechanism but ignored the delay aspect of performances. A frame delay analysis based on a similar Markov model is introduced in [10], which only covers the mean service time under the network saturation condition. Access delays of slotted ALOHA and CSMA without considering scheduling are given in [11]-[13]. It is arduous to study the queuing delay without a full analysis of the service time distribution that incorporates the scheduling algorithm. In facing the difficulty of precisely modeling the system, many simulation results on the performance of CSMA were reported in [14]-[16]. In a nutshell, any models of the network that ignored the queuing behavior of the input buffer of each node were unable to completely tackle the stability and delay issues of the protocol.

The intent of this paper is to provide a complete analysis of NP-CSMA/CA based on a multi-queue single-server model of the multiple access channel with $n$ nodes, as shown in Fig. 1, that are governed by NP-CSMA/CA protocol. In this two-stage queuing model as mentioned in [17], the input buffer of each node in the first stage is a FIFO (first-in-first-out) queue, and we assume that the input to each node is a Bernoulli process with rate $\lambda$. The second stage is a fictitious queue that consists of all HOL (head-of-line) packets of $n$ nodes with aggregate attempt rate $G$. According to the theorem on the superposition of point processes proved in [18], the aggregate attempts form a Poisson process for large population. The service time of HOL packets depends on the backoff scheduling algorithm when collision occurs. Therefore, it is impossible to achieve a comprehensive analysis of the system without considering the impact of the scheduling algorithm on the input queuing behavior. As mentioned earlier, most existing studies [8]-[11] and [19] focused on the aggregate attempts generated by HOL packets that input to the second-stage queue, from which an overall network throughput can be easily obtained. However, the stability and bounded delay conditions of each individual input queue cannot be comprehended by those simplified models. This point is clearly stated in [20] that focusing on aggregate attempt rate $G$ only is not enough to study the effect of retransmission policy.

The second-stage queue is characterized by a system model of the transmission channel, from which we derived the characteristic equation of the throughput. The result is consistent with that reported in [2] and [3]. We then model the first stage input buffer as a Geo/G/1 queue, in which the service time is specified by a Markov chain that represents the backoff scheduling algorithm. The complete queuing analysis enables us to specify the stable regions of both throughput and bounded mean delay. The network has a stable throughput when the channel's output rate equals the aggregate input rate. Our queuing analysis shows that the condition of throughput stability is determined by the first moment of service time, or offered load. Moreover, the Pollaczek-Khinchin (P-K) formula of Geo/G/1 queue indicates that a bounded mean delay also requires a bounded second moment of service time, which depends on the retransmission scheduling algorithm employed in the NP-CSMA protocol.

Two retransmission schemes are of importance, Geometric Retransmission and Exponential Backoff. For the Geometric Retransmission scheme, we find that the stable region of throughput is the same as the bounded delay region for a finite number of users. However, it is intrinsically unstable if the number of nodes in the network is infinite. For the Exponential Backoff scheme, there exists a throughput stable region even for an infinite population, which agrees with the result of [8] that the network with Exponential Backoff can have stable throughput even if the number of nodes $n \to \infty$. This scheme is more robust and can scale to large numbers of nodes.

The rest of this paper is organized as follows. The system model and input buffer model that give the characteristic throughput equation of the NP-CSMA protocol are presented in section II and section III, respectively. The general stable condition is described in section IV. The throughput and delay stability issues of the Geometric Retransmission and Exponential Backoff schemes are discussed in section V and VI, respectively. The conclusion is given in section VII.

## II. MARKOV CHAIN OF SLOTTED NP-CSMA

In a slotted NP-CSMA network, the time axis is slotted and the network is synchronized. Packets can be sent only at the beginning of a timeslot, and each packet transmission takes one slot time. We adopt the assumptions made in [2] that the processing time for sending an acknowledgement packet is negligible, and the ratio of propagation delay to packet transmission time is $a$. Then the timeslot can be further divided into mini-slots of slot size $a$. According to the channel status, the time axis can be considered as a sequence of alternating busy periods and idle periods as shown in Fig. 2a. The duration of a busy period is $1+a$ timeslots because no packets will attempt to transmit in the mini-slot immediately following a busy period.

The length of an idle period is a random variable with geometric distribution. In each busy period, either a packet is successfully transmitted or a collision occurred. Therefore, the channel will be in one of three states: idle (Idle), successful transmission (Suc), and collision (Col). The state transitions of the channel can be described by the Markov chain shown in Fig. 3.

The limiting probabilities $\pi_{Idle}, \pi_{Suc}$, and $\pi_{Col}$ of the Markov chain satisfy the set of equations:

$$\begin{cases} (1-P_{Idle,Idle}) \cdot \pi_{Idle} = P_{Suc,Idle} \cdot \pi_{Suc} + P_{Col,Idle} \cdot \pi_{Col} \\ (1-P_{Suc,Suc}) \cdot \pi_{Suc} = P_{Idle,Suc} \cdot \pi_{Idle} + P_{Col,Suc} \cdot \pi_{Col} \\ (1-P_{Col,Col}) \cdot \pi_{Col} = P_{Idle,Col} \cdot \pi_{Idle} + P_{Suc,Col} \cdot \pi_{Suc} \end{cases}. \quad (1)$$

Since packets attempt to access the channel only when the channel is detected as idle, the attempt rate in any busy period is zero. During an idle period, the aggregate attempts generated by all fresh and re-scheduled HOL packets form a Poisson stream with rate $G$, and the probability that no attempts is generated in a mini-slot is $e^{-aG}$. Viewed by the system, a transmission period is successful only if there is exactly one attempting packet in a mini-slot with the probability $aGe^{-aG}$. Hence, the transition probabilities of the Markov chain are given by

$$\begin{cases} P_{Idle,Idle} = P_{Suc,Idle} = P_{Col,Idle} = e^{-aG} \\ P_{Suc,Suc} = P_{Idle,Suc} = P_{Col,Suc} = aGe^{-aG} \\ P_{Col,Col} = P_{Idle,Col} = P_{Suc,Col} = 1 - e^{-aG} - aGe^{-aG} \end{cases}. \quad (2)$$

It is straightforward to show that the limiting probabilities are given as follows:

$$\pi_{Idle} = e^{-aG}, \quad \pi_{Suc} = aGe^{-aG} \text{ and } \pi_{Col} = 1 - e^{-aG} - aGe^{-aG}. \quad (3)$$

The time-average probabilities of each state can be easily obtained from sojourn times of Idle, Suc, and Col states $t_{Idle} = a$, $t_{Suc} = 1+a$, and $t_{Col} = 1+a$, respectively. Of particular importance, the probability that the channel is in Suc state is given by

$$\tilde{\pi}_{Suc} = \frac{(1+a)\pi_{Suc}}{(1+a)(\pi_{Suc} + \pi_{Col}) + a \cdot \pi_{Idle}} = \frac{(1+a)aGe^{-aG}}{1+a-e^{-aG}}. \quad (4)$$

Since the successful transmission of a packet only takes $1/(1+a)$ of the time in a busy period, the throughput of the network is defined by the fraction of the time that the channel is productive:

$$\hat{\lambda} = \frac{\tilde{\pi}_{Suc}}{(1+a)} = \frac{aGe^{-aG}}{1+a-e^{-aG}}, \quad (5)$$

which is consistent with Kleinrock and Tobagi's result [2]. Although the network throughput can be obtained from the above model, it is by no means a comprehensive performance analysis of the system, because this model does not consider the re-scheduling of collided HOL packets. The next section describes a more detailed queuing model of the input buffer in which the service time distribution is derived in cooperation with the backoff scheduling scheme.

### III. QUEUING MODEL OF INPUT BUFFER

The input buffer of each node is modeled as a Geo/G/1 with Bernoulli arrival process of rate $\lambda$ packets/timeslot. In the NP-CSMA protocol, we consider a general backoff scheme, called the *K*-Exponential Backoff algorithm, for contention resolutions. A fresh HOL packet is sent only when the idle channel is being sensed, and the packet is scheduled for retransmission at a later time if a collision occurred. A backlogged HOL packet is in phase $i$ if it has encountered collisions $i$ times. The *K*-Exponential Backoff algorithm allows an HOL packet in phase $i$ to retransmit with probability $q^i$, for $i = 1, \ldots, K$, where $0 < q < 1$ is the retransmission factor and $K$ is the cut-off phase. That is, the retransmission probability decreases exponentially with the number of collisions, up to $K$ times, experienced by the HOL packet.

Let $\alpha$ be the probability that the channel is being sensed in idle. Besides idle periods, the channel will also be sensed in idle state in the first mini-slot of each busy period because of the propagation delay. Therefore, the time-average probability that the channel is sensed idle can be calculated from the limiting probabilities given by (3) as follows:

$$\alpha = \frac{a \cdot \pi_{Idle} + a \cdot \pi_{Suc} + a \cdot \pi_{Col}}{a \cdot \pi_{Idle} + (1+a)(\pi_{Suc} + \pi_{Col})} = \frac{a}{1+a-e^{-aG}}. \tag{6}$$

As shown in Fig. 2b, the signal received by a node is shifted one mini-slot due to the propagation delay. Thus, the idle mini-slot immediately following a transmission period is still considered to be a part of the busy period by HOL packets. Also because of the propagation delay, a transmission can be successful only if all other nodes do not attempt to send in the first mini-slot of the busy period. Thus, the probability of a successful transmission of an HOL packet is $p = e^{-aG}$.

In the NP-CSMA protocol, an HOL packet in phase $i$ can be in one of the three states: sensing ($S_i$), transmission ($F_i$), and waiting ($W_i$), $i = 1, \ldots, K$ at any time. The state transition diagram is shown in Fig. 4. If an idle channel is detected in the sensing state $S_i$, then the packet will be transmitted in state $F_i$ with probability $q^i$. If a busy channel is sensed, the packet is in waiting state $W_i$. If the transmission is successful, a fresh HOL packet starts with the initial sensing state $S_0$; otherwise, the collided packet is moved to phase $i + 1$ and the above process repeats starting from the sensing state $S_{i+1}$. Moreover, we assume that a new packet arrived at an empty buffer will be delayed for one slot time in the waiting state if the channel is busy. The duration of this waiting time is not specified in non-persistent CSMA protocol. In fact, the sojourn time of the waiting state will not affect our analysis. Different waiting time may result in different delay, but the analyses of the stable throughput region and bounded delay region remain the same. Hence, the waiting time is assumed to be 1 slot time for the sake of simplicity.

From the Markov chain described in Fig. 4, we obtain the following set of state equations:

$$s_0 = w_0 + pf_0 + pf_1 + \cdots + pf_K, \quad w_0 = (1-\alpha)s_0, \quad f_0 = \alpha s_0;$$
$$s_i = \alpha(1-q^i)s_i + w_i + (1-p)f_{i-1}, \quad w_i = (1-\alpha)s_i, \quad f_i = \alpha q^i s_i, \qquad \text{for } i = 1,\ldots,K-1; \quad (7)$$
$$s_K = \alpha(1-q^K)s_K + w_K + (1-p)f_{K-1} + (1-p)f_K, \quad w_K = (1-\alpha)s_K, \quad f_K = \alpha q^K s_K.$$

It is easy to show from (7) that if $p + q > 1$ then all states of the Markov chain are positive recurrent and aperiodic. Let $s_i$, $f_i$, and $w_i$ be the limiting probabilities of states $S_i$, $F_i$, and $W_i$, respectively. Thus, the time-average probabilities of those states can be determined from (7) with the sojourn time $t_{S_i} = a$, $t_{F_i} = 1$, and $t_{W_i} = 1$ of states $S_i$, $F_i$, and $W_i$, respectively, for $i = 0, \ldots, K$ as follows:

$$\tilde{f}_i = \alpha p(p+q-1)(1-p)^i / D, \; i=0,1,\ldots,K-1; \quad \tilde{f}_K = \alpha(p+q-1)(1-p)^K / D;$$
$$\tilde{s}_i = ap(p+q-1)(1-p)^i / q^i D, \; i=0,1,\ldots,K-1; \quad \tilde{s}_K = a(p+q-1)(1-p)^K / q^K D; \qquad (8)$$
$$\tilde{w}_i = (1-\alpha)p(p+q-1)(1-p)^i / q^i D, \; i=0,1,\ldots,K-1; \quad \tilde{w}_K = (1-\alpha)(p+q-1)(1-p)^K / q^K D,$$

where $D = (1+a-\alpha)\left[pq - (1-q)(1-p)(1-p)^K / q^K\right] + \alpha(p+q-1)$.

The offered load $\rho$ of each input queue is the probability that the queue is non-empty. It is the basic measure for analyzing the performance of each input buffer. The input rate $\lambda$ of the Bernoulli arrival process can be interpreted as the probability of finding a packet arrived at input in any time slot. Each input packet will eventually become a fresh HOL packet, and visit the transmission state in phase 0 for one slot time. Therefore, the input rate $\lambda$ should equal $\rho \tilde{f}_0$, the probability of finding an HOL packet in the phase 0 transmission state in any time slot. With the time-average probability of state $F_0$ given in (8), the expression of the offered load can be obtained as follows:

$$\rho = \frac{\lambda}{\tilde{f}_0} = \frac{\lambda(1+a-\alpha)}{\alpha p(p+q-1)} \left[ pq - (1-q)(1-p)\left(\frac{1-p}{q}\right)^K \right] + \frac{\lambda}{p}. \quad (9)$$

Hence, the mean service time of each HOL packet $E[X]$ is equal to $1/\tilde{f}_0$. We show in the next theorem that the network throughput derived from the service time of HOL packet is the same as (5), which was obtained from the system model. Although the service time is obviously dependent on the retransmission factor $q$ of the backoff scheduling, the fact that the throughput is invariant with respect to $q$ implies that the stability of the system cannot be determined by the characteristic equation of throughput alone; it is mainly related to the queuing behavior of each input buffer.

**Theorem 1.** *For buffered non-persistent CSMA with K-Exponential Backoff, the throughput in equilibrium is given by*

$$\hat{\lambda}_{out} = \frac{-p \ln p}{1+a-p} = \frac{aGe^{-aG}}{1+a-e^{-aG}}. \quad (10)$$

Proof: For NP-CSMA protocol, a node is ready to send an HOL packet only if an idle channel has been detected. The probability of successful transmission from a desired node is the conditional probability that none of other nodes access the channel given that all nodes sense the channel idle,

$$\begin{aligned} p &= \Pr\{\text{none of other } n-1 \text{ nodes access the channel} \mid \text{channel is sensed idle}\} \\ &= \frac{\Pr\{\text{none of other } n-1 \text{ nodes access the channel}\}}{\Pr\{\text{channel is sensed idle}\}}. \end{aligned} \quad (11)$$

If no one access the channel, it means that all other $n-1$ nodes are either empty, or in sensing states but not accessing the channel. Thus, we have

Pr{none of other $n-1$ nodes access the channel}

$$= [\Pr\{\text{node is empty}\} + \Pr\{\text{node is in sensing but not scheduled to be sent}\}]^{n-1} \quad (12)$$

$$= \left[(1-\rho) + \rho \sum_{i=0}^{K} \tilde{s}_i (1-q^i)\right]^{n-1} \overset{\text{for large } n}{=} \exp\left\{-n\rho\left[1 - \sum_{i=0}^{K} \tilde{s}_i (1-q^i)\right]\right\}.$$

The probability that the node senses an idle channel is given by

$$\Pr\{\text{channel is sensed idle}\} = [\Pr\{\text{node is empty}\} + \Pr\{\text{node is in sensing state}\}]^{n-1}$$

$$= \left[(1-\rho) + \rho \sum_{i=0}^{K} \tilde{s}_i\right]^{n-1} \overset{\text{for large } n}{=} \exp\left\{-n\rho\left[1 - \sum_{i=0}^{K} \tilde{s}_i\right]\right\}. \quad (13)$$

Substituting (6) and (9) into (12) and (13), then the probability of successful transmission $p$ defined by (11) can be expressed as

$$p = \exp\left(-\frac{a\hat{\lambda}}{\alpha p}\right) = \exp-\frac{\hat{\lambda}(1+a-p)}{p}. \quad (14)$$

It is easy to show from (14) that the network throughput in equilibrium is given by (10). □

It should be noted that the throughput given by (10) also agrees with that obtained by Kleinrock and Tobagi in [2] and our channel model in section II. The consistency between these approaches indicates that it is appropriate to adopt the independence assumption among input buffers. This is because the correlation among $n$ input queues becomes weak in a multi-queue system when $n$ is large [21].

Furthermore, let random variables $S_i^*$, $F_i^*$ and $W_i^*$ be the service completion time of an HOL packet, starting from the states $S_i$, $F_i$, and $W_i$, respectively, until it is successfully transmitted. We assume, without loss of generality, that $M = 1/a$ is an integer. It is easy to show from the Markov chain of Fig. 4 that the generating functions $S_i(z)$, $F_i(z)$, and $W_i(z)$ of these service completion times can be found by solving the following set of equations:

$$\begin{cases} S_i(z) = E\left[z^{S_i^*}\right] = \alpha(1-q^i)zS_i(z) + (1-\alpha)zW_i(z) + \alpha q^i zF_i(z), & \text{for } i = 0,1,...,K; \\ W_i(z) = E\left[z^{W_i^*}\right] = z^M S_i(z), & \text{for } i = 0,1,...,K; \\ F_i(z) = E\left[z^{F_i^*}\right] = pz^M + (1-p)z^M S_{i+1}(z), & \text{for } i = 0,1,...,K-1; \\ F_K(z) = F_{K-1}(z). \end{cases} \quad (15)$$

Let $X$ denote the service time of a HOL packet. Since the service of each HOL packet starts from state $S_0$, therefore the first and second moments of service time, $E[X]$ and $E[X^2]$, can be derived from the generating function $S_0(z)$, which is given in Appendix I. Note that the mean service time derived in Appendix I is consistent with the expression $1/\tilde{f}_0$ given by (9). Based on the queuing model of the input buffer, we will investigate various stability issues of NP-CSMA concerning throughput and delay in the following sections.

## IV. STABLE REGIONS

Intuitively, if the attempt rate $G$ is too high, the probability of successful transmission $p = e^{-aG}$ will be too small; if it is too low, then the channel may not be fully utilized. Similar to the Aloha system discussed in [17] that the maximum throughput $\hat{\lambda}_{max}$ can be expressed in terms of the Lambert W function $W_0(z)$, which was first considered by J. Lambert in 1758, and later studied by L. Euler [22]. Taking the differentiation of the throughput equation (10), the maximum throughput $\hat{\lambda}_{max}$ can be derived with a corresponding attempt rate $G^*$ and expressed as below,

$$\hat{\lambda}_{max} = \frac{\left[W_0\left(-\frac{e^{-1}}{1+a}\right)+1\right]\exp\left[-W_0\left(-\frac{e^{-1}}{1+a}\right)-1\right]}{1+a-\exp\left[-W_0\left(-\frac{e^{-1}}{1+a}\right)-1\right]}. \tag{16}$$

As plotted in Fig. 5, for any attempt rate $G < G^*$, the network throughput increases with attempt rate $G$; otherwise, the throughput decreases with $G$. Therefore, for any throughput smaller than the maximum throughput, $\hat{\lambda} < \hat{\lambda}_{max}$, the characteristic equation (10) has two roots; the smaller and larger roots are denoted as $G_S(\hat{\lambda})$ and $G_L(\hat{\lambda})$, respectively. Considering the tradeoff between $G$ and $p$, we know the attempt rate $G$ should be bounded in the range between $G_S(\hat{\lambda})$ and $G_L(\hat{\lambda})$ to ensure a stable network throughput $\hat{\lambda}_{out} = \hat{\lambda}$. This point is illustrated by an example shown in Fig. 5, in which the network has a stable throughput of $\hat{\lambda}_{out} = \hat{\lambda} = 0.3$ when $G$ is within the range [$G_S \approx 0.5$, $G_L \approx 18.9$]. Specifically, a necessary condition of stable throughput of the entire system can be stated as follows:

**Stable Throughput Condition:**

**STC.** *For any input rate $\hat{\lambda} < \hat{\lambda}_{max}$, the attempt rate G should satisfy*

$$G_S(\hat{\lambda}) \leq G \leq G_L(\hat{\lambda}). \tag{17}$$

In general, the attempt rate $G$ is an implicit function of the retransmission factor $q$ associated with the underlying scheduling algorithm. This functional relationship, together with the inequality (17), determines a stable region of $q$. For the NP-CSMA protocol with exponential backoff scheduling algorithm, the expression of attempt rate $G$ in terms of $q$ is given below.

Suppose that there are a total $n_b = \sum_{i=1}^{K} n_i$ backlogged HOL packets in an idle period, in which $n_i$ packets are in the sensing state of phase $i$, for $i = 1,\ldots, K$. With the Bernoulli arrival process of rate $\lambda$, an empty node sends a packet if there is a new arrival with probability $\lambda$; and a backlogged node with an HOL packet in phase $i$ will send the HOL packet with probability $q^i$. Hence, the attempt rate $G$ is

$$G = (n - n_b)\lambda + \sum_{i=1}^{K} n_i q^i. \tag{18}$$

Let $\phi_i = \tilde{s}_i / \sum_{i=0}^{K} \tilde{s}_i$ be the probability that an HOL packet is in sensing state of phase $i$, for $i = 0,1,\ldots, K$. The attempt rate $G$ can be obtained from (8) and (18), and expressed as follows:

$$G = n\phi_0 \lambda + \sum_{i=1}^{K} n\phi_i q^i = \frac{\hat{\lambda}\tilde{s}_0 + n\sum_{i=1}^{K}\tilde{s}_i q^i}{\sum_{i=0}^{K}\tilde{s}_i} = \frac{[n(1-p) + \hat{\lambda}p](p+q-1)}{qp + (p+q-1-qp)\left(\frac{1-p}{q}\right)^K}. \tag{19}$$

The retransmission factor $q$ can be expressed as a function of the attempt rate $G$, denoted as $q = h(G)$, by solving equation (19). It is easy to show that the function $h(G)$ is monotonic increasing with respect to $G$. Thus, a region of the retransmission factor $q$, called *stable throughput region I*, can be specified by the combination of (17) and (19) as follows:

$$q \in R_I = [h(G_S), h(G_L)]. \tag{20}$$

The network throughput is defined by $\hat{\lambda}_{out} = \min\{n\tilde{f}_0, \hat{\lambda}\}$ and the stable throughput condition

**STC** ensures that $\hat{\lambda}_{out} = \hat{\lambda}$. It follows that the **STC** implies $n\tilde{f}_0 \geq \hat{\lambda} = n\lambda$ which means the offered load $\rho \leq 1$. On the other hand, it can be shown from (9) that the offered load $\rho$ is a monotonic increasing function of the retransmission factor q if the attempt rate is bounded in the range $G_S \leq G \leq G_L$, or equivalently $q \in R_I = [h(G_S), h(G_L)]$. In particular, the attempt rate G will reach $G_L$ when the offered load $=1$. Therefore, if the retransmission factor q is chosen from the stable region $R_I$, the offered load $\rho$ of Geo/G/1 queue of each input buffer should be strictly less than 1, which is the stable condition of any queuing systems that the arrival rate should be less than the service rate.

Let $G_t$ be the instantaneous attempt rate in timeslot $t$. Since the steady state mean attempt rate $G$ of a Poisson stream is defined by the time average of instantaneous attempt rates:

$$G = \lim_{T \to \infty} \frac{1}{T} \sum_{t=0}^{T} G_t . \tag{21}$$

Thus, the stable throughput region I, determined by the steady state mean attempt rate $G$, could be too loose to cope with the fluctuations of the instantaneous attempt rates $G_t$, and the network may become unstable if $G_t$ is larger than $G_L$ at some point in time $t$. To prevent this risk, the following conservative estimate $\hat{G}_L$ of the upper bound of the stable throughput region is given in (54) of Appendix II:

$$G_L = \hat{G}_L + 3\sqrt{\hat{G}_L \left(1 - \hat{G}_L / n\right)} . \tag{22}$$

which takes the variance of $G_t$ into consideration to ensure that the instantaneous attempt rate $G_t$ will be strictly less than $G_L$ with a probability higher than .99 at any instants of time. Solving (54) for $\hat{G}_L$, we have

$$\hat{G}_L = \frac{2nG_L + 9n - 3\sqrt{4n^2 G_L + 9n^2 - 4nG_L^2}}{2n + 18} . \tag{23}$$

On the other hand, the risk is marginal if the instantaneous attempt rate $G_t$ drops below $G_S$ at some time $t$. Because the probability of success $p$ becomes large when the attempt rate is so low, the throughput eventually climbs back and becomes equal to the input rate $\hat{\lambda}$ in steady state.

Therefore, a conservative region of retransmission factor $q$, called *stable throughput region II*, can be specified by (23) as follows:

$$R_{II} = \left[ h(G_S), h(\hat{G}_L) \right]. \tag{24}$$

As we mentioned before, the throughput stable condition (17) is not sufficient to guarantee a bounded mean delay of packets queued in each input buffer. To complete the analysis of stable regions, we deduce the following additional constraint from our queuing model of the input buffer.

**Bounded Delay Condition:**

**BDC.** *The Pollaczek-Kenichi formula for mean delay $E[T]$ of Geo/G/1 queue* [21]

$$E[T] = E[X] + \frac{\lambda E[X^2]}{2(1 - \lambda E[X])} \tag{25}$$

*requires bounded second moment of service time* $0 < E[X^2] < \infty$.

The condition **BDC** is more restrictive than the condition **STC**. A region of the factor $q$, denoted as $R_D$, that guarantees bounded mean delay can be determined by the second moment of service time $E[X^2]$. In general, the bounded delay region is a subset of the throughput stable region. Detailed discussions on these stable regions of two particularly important scheduling schemes, Geometric Retransmission and Exponential Backoff, are provided below.

## V. ANALYSIS OF GEOMETRIC RETRANSMISSION

The simplest exponential backoff scheduling is the Geometric Retransmission with cutoff phase $K=1$. In this section, we study both stable throughput region and bounded delay region of the retransmission factor $q$ of Geometric Retransmission.

### A. *Stable Throughput Condition of Geometric Retransmission*

The attempt rate $G$ for this scheme can be obtained by substituting $K = 1$ and $p = e^{-aG}$ into (19), which yields the following expression:

$$G = \frac{\hat{\lambda} pq + n(1-p)q}{pq + 1 - p}, \tag{26}$$

or, equivalently, the retransmission factor $q$ can be expressed as a function of $G$ given by

$$q = h(G) = \frac{G}{n + (\hat{\lambda} - G)e^{-aG} / (1 - e^{-aG})}. \quad (27)$$

For retransmission factor $q$ in the range $0 < q < 1$, it is easy to show that the function $h(G)$ is monotonic increasing with respect to the attempt rate $G$. Hence, the following stable throughput region I for Geometric Retransmission can be immediately obtained from (20) and (27):

$$R_I^{Ge} = [h(G_S), h(G_L)] = \left[ \frac{G_S}{n + (\hat{\lambda} - G_S)e^{-aG_S} / (1 - e^{-aG_S})}, \frac{G_L}{n + (\hat{\lambda} - G_L)e^{-aG_L} / (1 - e^{-aG_L})} \right]. \quad (28)$$

For $a = 0.1$ and $n = 50$, the stable region (28) of retransmission factor $q$ lies in the shaded part displayed in Fig. 6, which shows the stable throughput range of $q$ for a given input rate $\hat{\lambda} \leq \hat{\lambda}_{max}$.

Moreover, the conservative stable throughput region $R_{II}^{Ge}$ for Geometric Retransmission can be derived from (24) and (27), and given as follows:

$$R_{II}^{Ge} = \left[ h(G_S), h(\hat{G}_L) \right] = \left[ \frac{G_S}{n + (\hat{\lambda} - G_S)e^{-aG_S} / (1 - e^{-aG_S})}, \frac{\hat{G}_L}{n + (\hat{\lambda} - \hat{G}_L)e^{-a\hat{G}_L} / (1 - e^{-a\hat{G}_L})} \right]. \quad (29)$$

The above risk prevention mechanism is achieved at the expense of sacrificing the maximum throughput $\hat{\lambda}_{max}$. In the stable throughput region $R_{II}^{Ge}$, only a smaller maximum throughput $\hat{\lambda}_{max}^{Ge} < \hat{\lambda}_{max}$ can be achieved when $G_S(\hat{\lambda}_{max}^{Ge}) = \hat{G}_L(\hat{\lambda}_{max}^{Ge})$, as illustrated in Fig. 6.

For Geometric Retransmission, both throughput stable regions I (28) and II (29) will be vanished when the number of nodes $n$ goes to infinity; hence, the network is unstable for any retransmission factor $q$. When the network is fully congested, the attempt rate $G$ for a large number of nodes $n$ is given by (27) as follows:

$$G = nq. \quad (30)$$

That is, a large number of packets will be backlogged in phase 1 when $n \to \infty$. As a result, the probability of successful transmission $p = e^{-anq} \to 0$, and the network will become unstable. For the same reason, Exponential Backoff with finite cut-off phase $K$ is not scalable. The instability of Geometric Retransmission has also been reported by others in [8] and [19].

Fig. 7 shows that our analysis is consistent with simulation results for a network with $n = 50$ and $a = 0.1$. Both results indicate that a stable throughput can be achieved if $q$ is chosen within the region $R_{II}^{Ge}$, and the throughput quickly drops to zero outside this region.

B. *Bounded Delay Condition of Geometric Retransmission*

For the Geometric Retransmission scheme, the following second moment of service time is derived through (42) of Appendix I as follows:

$$E[X^2] = \frac{a^2 A(p,q)}{\alpha^2 p^2 q^2}, \qquad (31)$$

where $A(p, q)$ is a polynomial of $p$ and $q$, and $A(p, q) > 0$. From the expression shown in Appendix I, the second moment is always bounded within the stable throughput region II. Therefore, the bounded delay region $R_D^{Ge}$ is the same as $R_{II}^{Ge}$ given by (29). According to Little's Law, the mean queueing delay is linearly related to the number of packets in the system. The simulation results for $n = 50$, $a = 0.1$, and $\hat{\lambda} = 0.3$ is shown in Fig. 8. Both analysis and simulation demonstrate that the mean delay is unbounded when $q$ is larger than 0.233, which is the upper bound of $R_{II}^{Ge}$ calculated from (29). In sum, our results indicate that the NP-CSMA protocol with Geometric Retransmission cannot be scaled to a large number of nodes $n$. A much more robust scheduling algorithm, Exponential Backoff, is discussed in the next section.

VI. ANALYSIS OF EXPONENTIAL BACKOFF

The Exponential Backoff scheme has been studied in many previous papers [7]-[11]. In this section, we discuss the stability and delay performance of this scheduling algorithm based on the stable throughput and bounded delay conditions specified in section IV.

A. *Stable throughput Condition of Exponential Backoff*

The following attempt rate $G$ of Exponential Backoff can be derived by taking the limit of (19) as $K$ goes infinity and then substituting $p = e^{-aG}$ into the resulting (19), which gives the following expression:

$$G = \left[n(1-p) + \hat{\lambda}p\right]\left(\frac{p+q-1}{pq}\right). \tag{32}$$

or, equivalently, the retransmission factor $q$ can be formulated as a function of attempt rate $G$ as

$$q = h(G) = \frac{1 - e^{-aG}}{1 - \frac{Ge^{-aG}}{n - (n-\hat{\lambda})e^{-aG}}}. \tag{33}$$

Again, this function $q = h(G)$ is monotonically increasing with respect to the attempt rate $G$. Hence, the following stable throughput region I of Exponential Backoff can be obtained from (20) and (33):

$$R_I^{Ex} = [h(G_S), h(G_L)] = \left[\frac{1 - e^{-aG_S}}{1 - G_S e^{-aG_S} / \left[n - (n-\hat{\lambda})e^{-aG_S}\right]}, \frac{1 - e^{-aG_L}}{1 - G_L e^{-aG_L} / \left[n - (n-\hat{\lambda})e^{-aG_L}\right]}\right]. \tag{34}$$

Similar to the Geometric retransmission scheme, from (24) and (33), a more conservative stable throughput region II for Exponential Backoff is given as follows:

$$R_{II}^{Ex} = [h(G_S), h(\hat{G}_L)] = \left[\frac{1 - e^{-aG_S}}{1 - G_S e^{-aG_S} / \left[n - (n-\hat{\lambda})e^{-aG_S}\right]}, \frac{1 - e^{-a\hat{G}_L}}{1 - \hat{G}_L e^{-a\hat{G}_L} / \left[n - (n-\hat{\lambda})e^{-a\hat{G}_L}\right]}\right]. \tag{35}$$

The retransmission factor $q \in R_{II}^{Ex}$ ensures that the instantaneous attempt rate $G_t$ is strictly less than the desired upper bound $G_L$ with a probability higher than .99 at any time $t$.

Fig. 9 shows various regions of Exponential Backoff with $n = 50$ and $a = 0.1$. The lower bound $h(G_S)$ and upper bound $h(G_L)$ of the stable throughput region $R_I^{Ex}$ that constitutes the upper curve are calculated from (34). Similar to the Geometric Retransmission scheme, the conservative stable throughput region $R_{II}^{Ex}$ derived from (35) is a subset of the region $R_I^{Ex}$. Thus, only a smaller maximum throughput $\hat{\lambda}_{max}^{Ex} < \hat{\lambda}_{max}$ can be achieved in the region $R_{II}^{Ex}$ when $G_S(\hat{\lambda}_{max}^{Ex}) = \hat{G}_L(\hat{\lambda}_{max}^{Ex})$. Our stability analysis is confirmed by the simulation results shown in Fig. 10, which depicts that a stable throughput can always be achieved if the retransmission factor $q$ is properly chosen from the stable throughput region $R_{II}^{Ex}$.

In contrast to Geometric Transmission, the following non-empty stable throughput region can be obtained from (35) when the number of nodes *n* goes to infinity:

$$R_{II}^{Ex} = [1-e^{-aG_S}, 1-e^{-a\hat{G}_L}], \qquad (36)$$

which coincides with Song's results on Aloha system with Exponential backoff shown in [8], [11] and [19] that the network throughput can be non-zero even when the number of nodes goes to infinity.

B. *Bounded Delay Condition of Exponential Backoff*

The second moment of service time of the Exponential Backoff scheme is given by (45) in Appendix I as follows:

$$E[X^2] = a^2 B(p,q) \lim_{K \to \infty} \left(\frac{1-p}{q^2}\right)^{K-1} + a^2 C(p,q), \qquad (37)$$

where *B(p,q)* and *C(p,q)* are two polynomials given by (46) and (47), respectively, in Appendix I. Thus, the retransmission factor should satisfy following condition to guarantee a bounded second moment of service time $E[X^2]$:

$$q^2 > 1-p, \qquad (38)$$

or equivalently,

$$q > \sqrt{1-e^{-aG}}. \qquad (39)$$

Since bounded delay implies stable throughput, thus the bounded delay region of Exponential Backoff can be specified by the combination of condition (39) and the stable throughput region $R_{II}^{Ex}$, and given as follows:

$$R_D^{Ex} = \left[\sqrt{1-e^{-aG_S}}, \frac{1-e^{-a\hat{G}_L}}{1-\hat{G}_L e^{-a\hat{G}_L}/\left[n-(n-\hat{\lambda})e^{-a\hat{G}_L}\right]}\right]. \qquad (40)$$

As shown in Fig. 9, the shaded bounded delay region $R_D^{Ex}$ is a subset of the stable throughput region $R_{II}^{Ex}$. Outside this bounded delay region when $q \in R_{II}^{Ex} \setminus R_D^{Ex}$, the system may still have a stable throughput but the mean delay will quickly jump up to an unacceptable level as shown in Fig. 11, which exhibits the number of backlogged packets in the entire system versus retransmission factor $q$ for a fixed aggregate input rate $\hat{\lambda} = 0.3$. It can be clearly seen that if 0.21<$q$<0.73, within the region of bounded delay, there is nearly zero backlogged packets in the system, and the mean delay is bounded according to Little's law in queuing theory. However, if the retransmission factor $q$ is greater than the upper bound of the delay region, the number of packets in the system suddenly becomes much larger even when $q$ is still within the throughput stable region, i.e., $q$>0.053. On the other hand, the simulation result of Fig. 11 also shows that the analytical lower bound of the region $R_D^{Ex}$ given in (40) is much more conservative than the upper bound. Furthermore, compared with Geometric Retransmission for the case when $n$ = 50 shown in Fig. 8, the Exponential Backoff has a larger bounded delay region for retransmission factor $q$.

In sum, the Exponential Backoff has two types of stable regions as shown in Fig. 9. The first type is the bounded delay region $R_D^{Ex}$, represented by the shaded area in Fig. 9, in which the network throughput is stable and the mean delay is bounded. The second type is the region $R_{II}^{Ex} \setminus R_D^{Ex}$ that only guarantees stable throughput but with unacceptable large mean delay. In the Exponential Backoff, it is possible that predominating backlogged packets are pushed to deep phases with very low retransmission probabilities when the network becomes congested. If a node tries to send its HOL packet, the successful probability will be very high. Once the backlogged HOL packet is cleared, then the channel may be "captured" by subsequent packets in the input buffer, which are all in phase 0, until the queue is cleared. During this time, it appears that the network throughput is still stable, but the variance of the service time of each individual packet can be infinitely large due to this unfairness of services caused by the capture effect mentioned in [23]. Nevertheless, compared with the Geometric Retransmission, the Exponential Backoff is more robust and can scale to a larger population.

## VII. CONCLUSION

A multi-queue-single-server model is proposed in this paper to explore the stability and delay issues of multiple access networks with the NP-CSMA protocol. We studied the throughput of the entire system and the performance of each individual input buffer. The *K*-Exponential Backoff scheduling algorithm is considered in the analysis of the service time of HOL packets. Based on this model, we formulated conditions on stable throughput and bounded mean delay according to the basic principle of queuing theory. Throughput stable region and bounded delay region of the retransmission factor are established for both the Geometric Retransmission and Exponential Backoff schemes. Our method is mainly based on the construction of Markov chains of the transmission channel and the service time of HOL packets; they can be easily generalized to investigate other MAC protocols, such as IEEE 802.11, in the future.

APPENDIX I.  SERVICE TIME DISTRIBUTION FOR NP-CSMA PROTOCOL

For the two backoff schemes of interest, the actual first moment $E[X]$ and second moment $E[X^2]$ of service time of HOL packets can be derived from (15) as follows:

**Geometric Retransmission (*K* = 1)**

$$E[X] = aS_0'(1) = \frac{a(1-p)(M+1+M\alpha q - M\alpha) + apq(M+1)}{\alpha pq} \quad (41)$$

$$E[X^2] = a^2\left(S_0''(1) + S_0'(1)\right) = \frac{a^2 A(p,q)}{\alpha^2 p^2 q^2}, \quad (42)$$

where 
$$A(p,q) = \alpha q(4M+p)(1-p)(1+M-\alpha M) + 2(1-p)\left[(1+M-\alpha M)^2 + (\alpha qM)^2\right] \quad (43)$$
$$+\alpha pq^2(1+M)(2M-\alpha M+p) + pq(2M-\alpha M+2)(pq+1-p)(1-\alpha)(1+M) + \alpha^2 pq^2 M.$$

**Exponential Backoff (*K* → ∞)**

$$E[X] = \frac{aM}{p} + \frac{aq(M+1-M\alpha)}{\alpha(p+q-1)} \quad (44)$$

$$E[X^2] = a^2 B(p,q) \lim_{K\to\infty}\left(\frac{1-p}{q^2}\right)^{K-1} + a^2 C(p,q), \quad (45)$$

where

$$B(p,q) = \frac{(p+q-1-2qp)(M+1-M\alpha)}{\alpha qp(q-1)(p+q-1)} + \frac{(M+1-M\alpha)q}{\alpha(q-1)(p+q^2-1)} - \frac{2(M+1-M\alpha)^2(1-p)}{\alpha^2(p+q^2-1)} \quad (46)$$

and

$$C(p,q) = 2M(1-p)\left(\frac{M}{p^2}-1\right) + \frac{M^2q(M-1)}{p^2(q-1)} + \frac{(1+M-2\alpha M)(M-1)[M+2p(q-1)]}{\alpha pq(q-1)(p+q-1)}$$
$$+ \frac{(M+1)[2(1-p)+Mq(1-\alpha)]}{\alpha(p+q-1)} + \frac{qM(M+1-M\alpha)[2p(1-p)(q-1)-M+1]}{\alpha p(q-1)(p+q-1)^2} \quad (47)$$
$$+ \frac{2(M+1-M\alpha)^2 q^2}{\alpha^2(p+q^2-1)} + S_0'(1).$$

APPENDIX II. VARIANCE OF INSTANTANEOUS ATTEMPT RATE $G$ FOR NP-CSMA PROTOCOL

For each node $i = 1, .., n$ at time $t$, let $X_i$ denote the binary random variable defined by: $X_i = 1$, if node $i$ attempts to transmit ; $X_i = 0$, otherwise. Let the mass function of $X_i$ be given by

$$P_{X_i}(x) = \begin{cases} 1-g & x=0 \\ g & x=1 \end{cases}. \quad (48)$$

The mean $\mu$ and variance $\sigma^2$ of $X_i$ are given as follows:

$$E[X_i] = \mu = g \quad \text{and} \quad E[(X_i - \mu)^2] = \sigma^2 = g(1-g). \quad (49)$$

Let $G_t = \sum_{i=1}^{n} X_i$ be the total number of attempts at time $t$. Then the mean and variance of $G_t$ are

$$E[G_t] = n\mu \quad \text{and} \quad Var[G_t] = n\sigma^2. \quad (50)$$

According to the Central Limit Theorem, the distribution of $G_t$ converges to the normal distribution when $n$ is large:

$$\lim_{n\to\infty} P\left(\frac{G_t - n\mu}{\sigma\sqrt{n}} \le z\right) = \lim_{n\to\infty} P\left(\frac{\sum_{i=1}^{n} X_i - n\mu}{\sigma\sqrt{n}} \le z\right) = \Phi(z). \quad (51)$$

Assuming that the system is ergodic, the estimate mean attempt rate $\hat{G}$ should equal the mean of $G_t$. It follows from (49) and (50) that the mean and variance of $G_t$ are given as

$$E[G_t] = n\mu = ng = \hat{G} \quad \text{and} \quad Var[G_t] = n\sigma^2 = \hat{G}\left(1 - \hat{G}/n\right). \quad (52)$$

According to the property of normal distribution, about 99% of the outcomes are within three standard deviations of the mean value. Therefore, we have

$$\lim_{n\to\infty} P\left(\frac{G_t - \hat{G}}{\sigma\sqrt{n}} \le 3\right) = \lim_{n\to\infty} P\left(\frac{\sum_{i=1}^{n} X_i - \hat{G}}{\sigma\sqrt{n}} \le 3\right) = \Phi(3) \approx 99\%. \tag{53}$$

To guarantee that the instantaneous attempt rate $G_t$ being bounded by $G_L$ at any time $t$, a conservative estimate $\hat{G}$ of the upper bound should satisfy the following:

$$G_L = \hat{G} + 3\sigma\sqrt{n} = \hat{G} + 3\sqrt{\hat{G}(1 - \hat{G}/n)}. \tag{54}$$

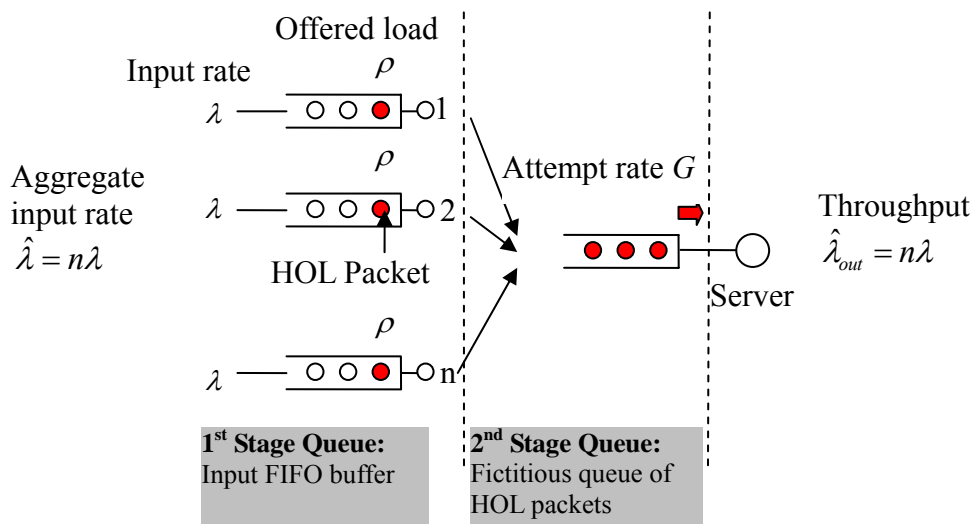

Fig. 1. A two-stage multi-queue single-server system.

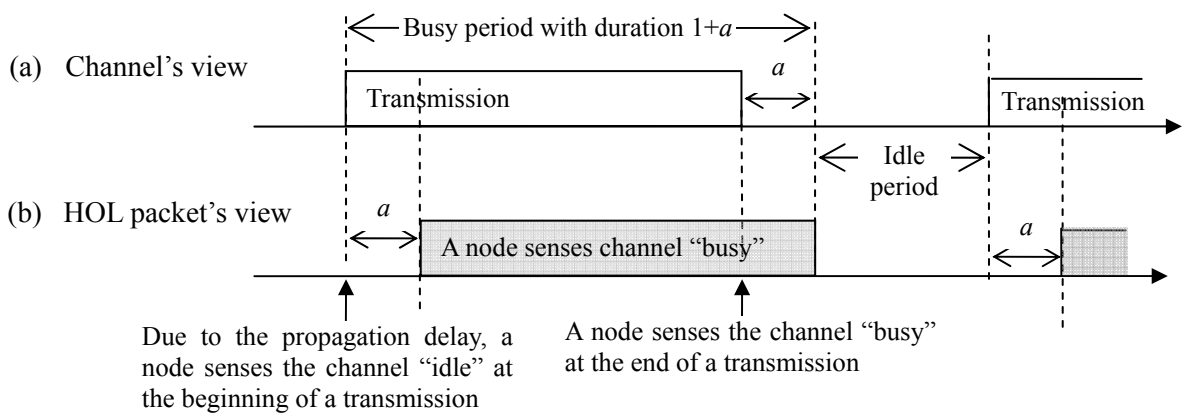

Fig. 2. Busy and idle periods of NP-CSMA protocol (a) viewed by channel and (b) viewed by HOL packets.

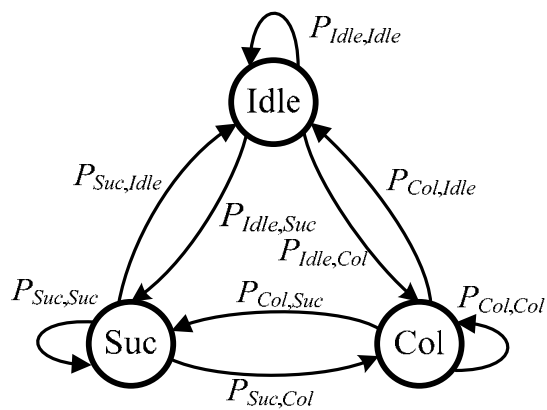

Fig. 3. Markov Chain of NP-CSMA channel.

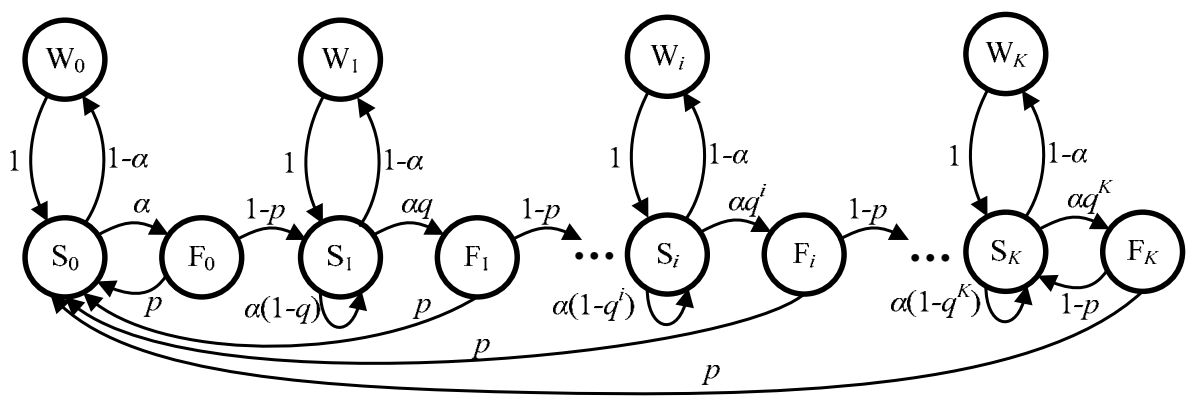

Fig. 4. Markov chain of HOL packet.

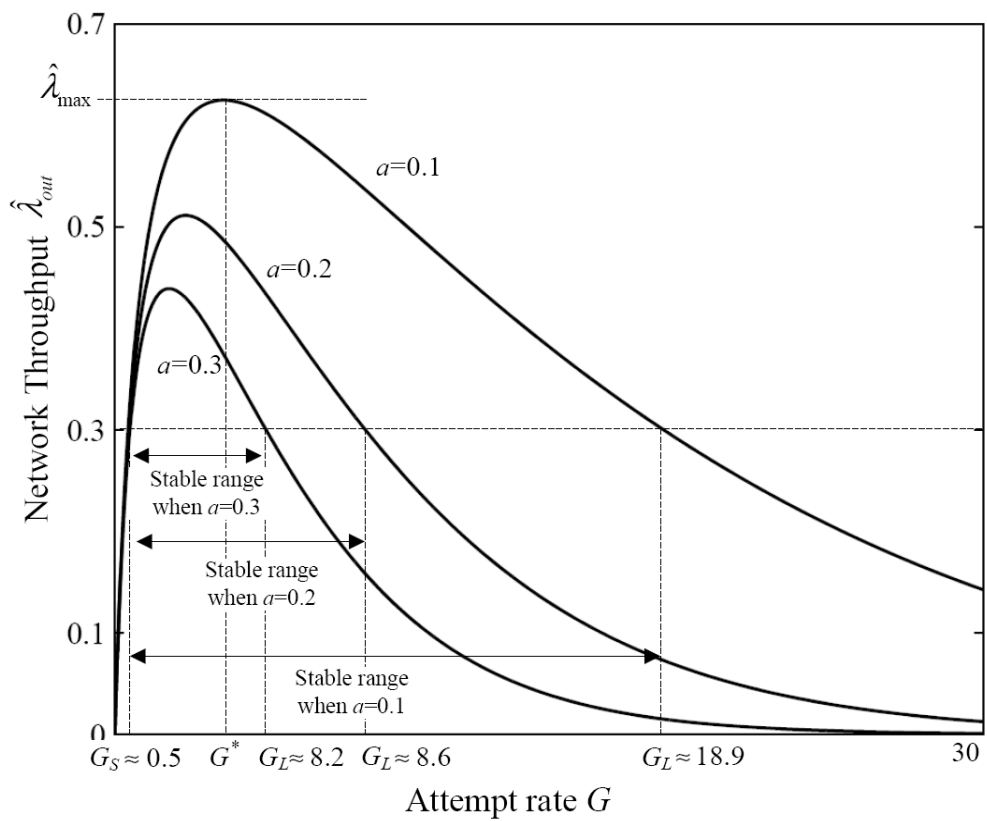

Fig. 5.    Throughput versus the attempt rate for NP-CSMA.

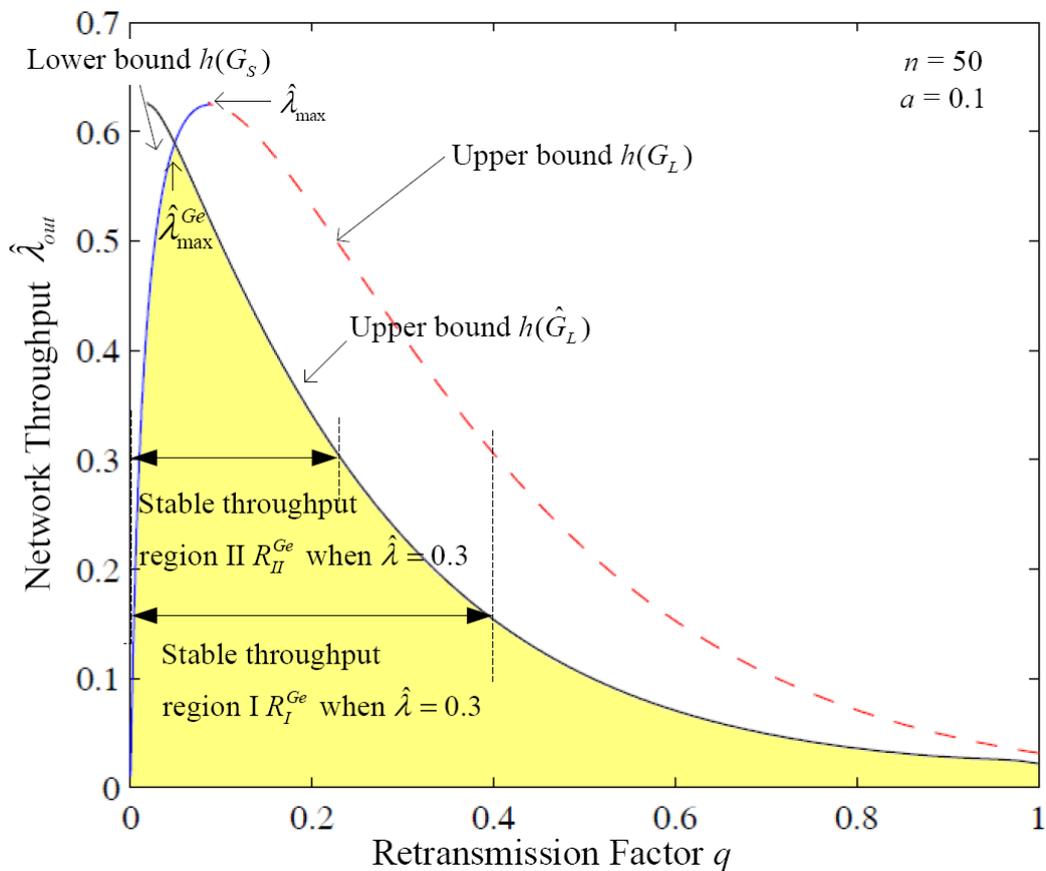

Fig. 6.    Stable throughput region $R_I^{Ge}$ and $R_{II}^{Ge}$ of Geometric Retransmission.

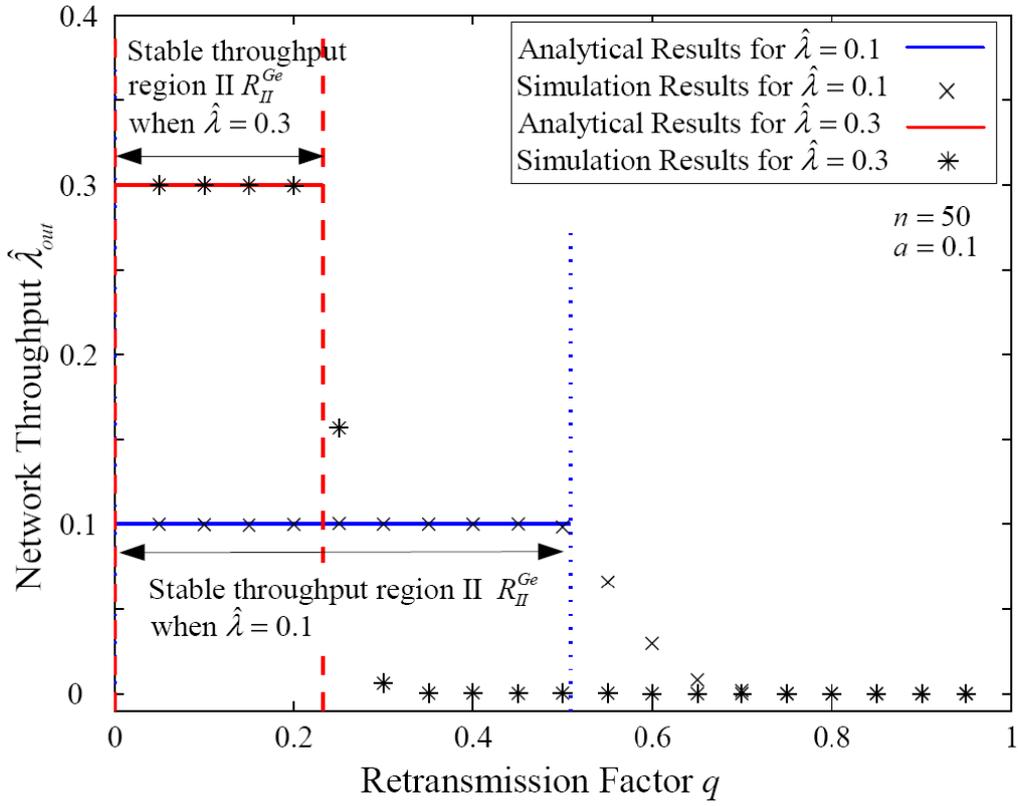

Fig. 7. Simulation results of stable throughput region II $R_{II}^{Ge}$.

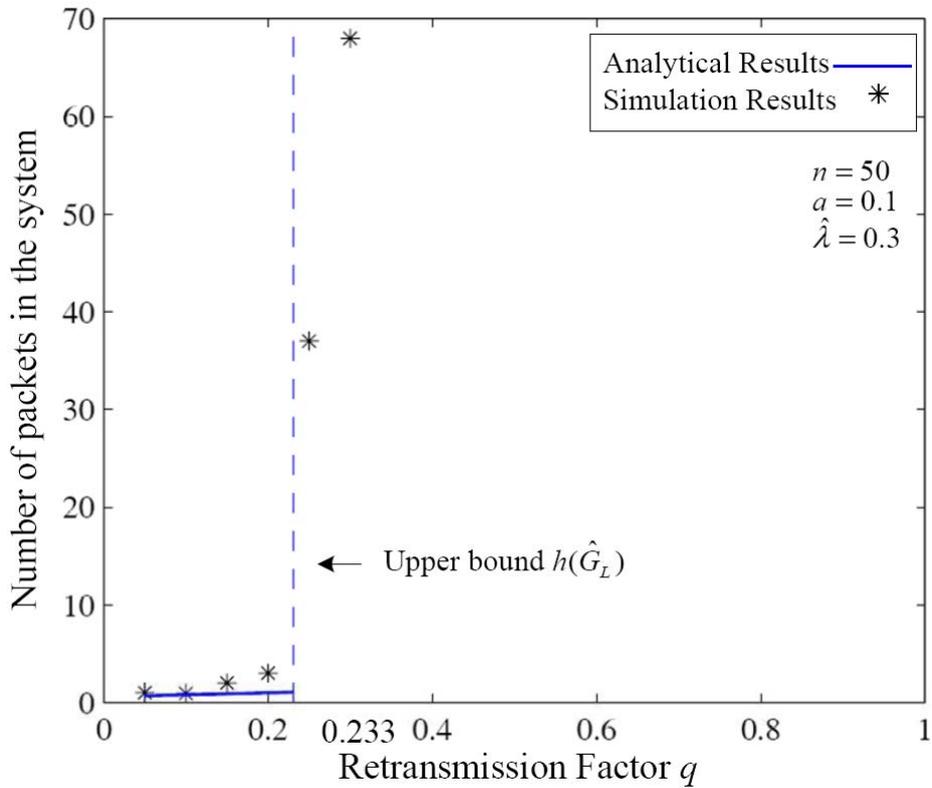

Fig. 8. Number of packets verses $q$ for Geometric Retransmission.

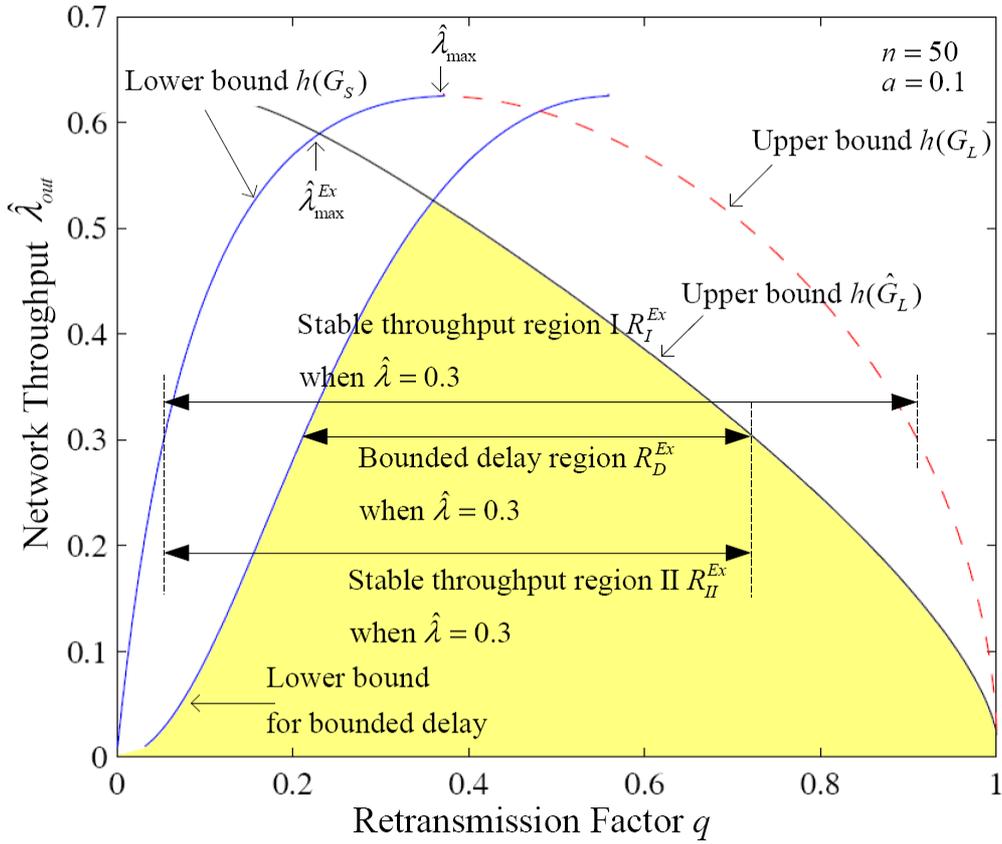

Fig. 9. Regions $R_I^{Ex}$, $R_{II}^{Ex}$ and $R_D^{Ex}$ of Exponential Backoff.

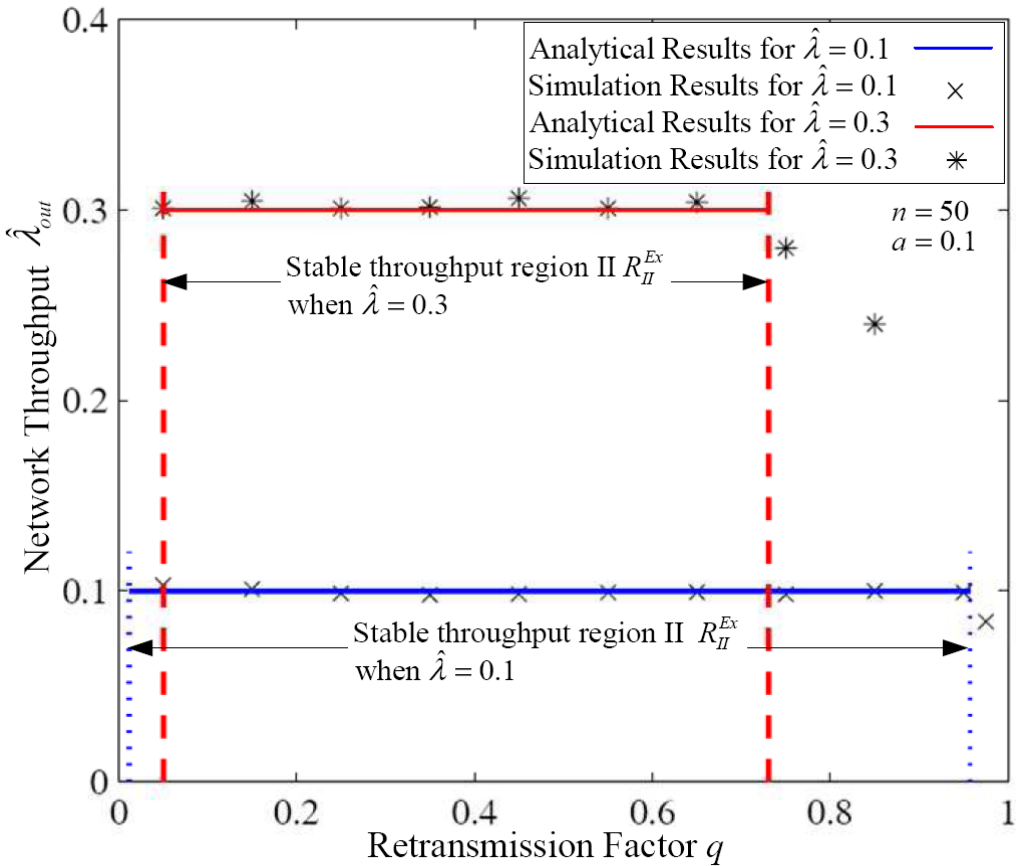

Fig. 10. Simulation results of stable throughput region II $R_{II}^{Ex}$.

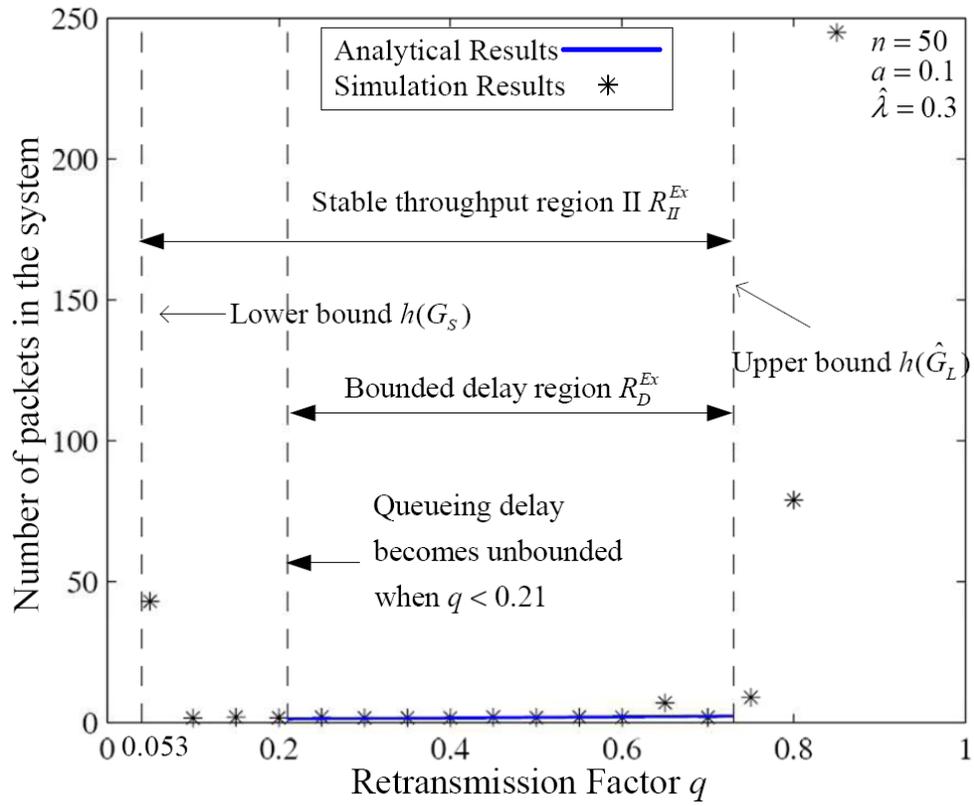

Fig. 11. Number of packet in the system verses $q$ for Exponential Backoff.